\documentclass[letter]{aa} 

\usepackage{graphicx}
\usepackage{txfonts}

\usepackage[T1]{fontenc}
\usepackage{nicefrac}
\usepackage[utf8]{inputenc}
\usepackage{amsmath}
\usepackage{xfrac}
\usepackage{mathtools}
\usepackage{booktabs}
\usepackage{multirow}
\usepackage[normalem]{ulem}

\usepackage{siunitx}
\DeclareSIUnit\parsec{pc}
\DeclareSIUnit\years{yr}
\DeclareSIUnit\Msol{M_{\odot}}
\DeclareSIUnit\Lsol{L_{\odot}}
\DeclareSIUnit\AU{au}
\DeclareSIUnit\om{\Omega}
\DeclareSIUnit\orb{T_{\mathrm{orb}}}
\DeclareSIUnit\scaleheight{H}
\newcommand{\rhod}{\rho_{\mathrm{d}}}

\newcommand{\amax}{a_{\mathrm{max}}}
\newcommand{\amin}{a_{\mathrm{min}}}

\newcommand{\St}{\mathrm{St}}

\newcommand{\OmK}{\Omega_\text{K}}

\makeatletter
\renewcommand*\aa@pageof{, page \thepage{} of \pageref*{LastPage}}
\makeatother

\definecolor{RoyalBlue}{HTML}{0071BC}
\definecolor{Maroon}{HTML}{AF3235}
\newcommand{\rev}[1]{{#1}}
\newcommand{\revII}[1]{{#1}}

\newcommand{\reportI}[1]{{#1}}

\usepackage{hyperref}
\usepackage{cleveref}

\begin{document}



   \title{\revII{Vertical shear instability with dust evolution and \\ consistent cooling times}}
   \subtitle{{On the importance of the initial dust distribution}}
   
   \author
   {
        Thomas Pfeil \inst{123}
        \and
        Til Birnstiel \inst{14}
        \and
        Hubert Klahr \inst{2}
    }

   \institute
   {
        University Observatory, Faculty of Physics, Ludwig-Maximilians-Universität München, Scheinerstr. 1, D-81679 Munich, Germany\\
        \email{tpfeil@usm.lmu.de}
        \and 
        Max-Planck-Institut f\"ur Astronomie, K\"onigstuhl 17, D-69117 Heidelberg, Germany
        \and  {Center for Computational Astrophysics, Flatiron Institute, 162 Fifth Avenue, New York, NY 10010, USA}
        \and Exzellenzcluster ORIGINS, Boltzmannstr. 2, D-85748 Garching, Germany
    }


 
  \abstract
   {Gas in protoplanetary disks mostly cools via thermal accommodation with dust particles. Thermal relaxation is thus highly sensitive to the local dust size distributions and the spatial distribution of the grains. \revII{So far, the interplay between thermal relaxation and gas turbulence has not been dynamically modeled in hydrodynamic simulations of protoplanetary disks with dust.}}
   {\revII{We aim to study the effects of the vertical shear instability (VSI) on the thermal relaxation times, and vice versa. We are particularly interested in the influence of the initial dust grain size on the VSI and whether the emerging turbulence is sustained over long timescales.}}
   {We ran three axisymmetric hydrodynamic simulations of a protoplanetary disk including four dust fluids that initially resemble MRN size distributions of different initial grain sizes. From the local dust densities, we calculated the thermal accommodation timescale of dust and gas and used the result as the thermal relaxation time of the gas in our simulation. We included the effect of dust growth by applying the monodisperse dust growth rate and the typical growth limits.}
   {We find that the emergence of the VSI is strongly dependent on the initial dust grain size. Coagulation also counteracts the emergence of hydrodynamic turbulence in our simulations, as shown by others before. Starting a simulation with larger grains (\SI{100}{\micro \meter}) generally leads to a less turbulent outcome. While the inner disk regions (within $\sim \SI{70}{au}$) develop turbulence in all three simulations, we find that the simulations with larger particles do not develop VSI in the outer disk.}
   {\revII{Our simulations with dynamically calculated thermal accommodation times based on the drifting and settling dust distribution} show that the VSI, once developed in a disk, can be sustained over long timescales, even if grain growth is occurring. The VSI corrugates the dust layer and even diffuses the smaller grains into the upper atmosphere, where they can cool the gas. Whether the instability can emerge \revII{for a specific stratification} depends on the initial dust grain sizes and the initial dust scale height. If the grains are initially $\gtrsim \SI{100}{\micro \meter}$ and if the level of turbulence is initially assumed to be low, we find \reportI{no VSI turbulence} in the outer disk regions.}


   \maketitle
%

\section{Introduction}
Thermal relaxation of temperature perturbations in {the gas of} protoplanetary disks occurs in a two-step process, where collisions thermally {accommodate} the slowly cooling gas with the quickly cooling dust grains, which then emit or absorb electromagnetic radiation \citep{Malygin2014, Woitke2015, Malygin2017, Barranco2018}. 
The local dust size distribution thus plays a crucial role when it comes to the determination of the thermal relaxation time of the gas. 
Many thermal and hydrodynamic instabilities depend on this timescale (e.g., convective overstability ,\cite{Klahr2014}; vertical shear instability, \cite{Urpin1998, Nelson2013}; zombie vortex instability, \cite{Marcus2015}). Turbulence resulting from these instabilities can redistribute the dust grains due to aerodynamic drag, altering the cooling timescale itself.
Continuous remodeling of the cooling times in hydrodynamic simulations of protoplanetary disks is thus desirable.

In this letter, we present axisymmetric simulations of protoplanetary disks that, for the first time, combine the effects of dust grain growth and the emergence of hydrodynamic turbulence on the cooling times. We are specifically interested in the effect on the vertical shear instability (VSI). Various studies have shown its strong dependence on the thermal relaxation time \citep[e.g.,][]{Urpin2003, Lin2015, Manger2021, Pfeil2021}.
We note that VSI requires rapid cooling on a timescale ideally shorter than 
\begin{equation}
\label{eq:tcrit}
    t_\mathrm{c}\lesssim \frac{H_\mathrm{g}}{R}\frac{|\beta_T|\gamma}{2(\gamma-1)}\left|\frac{z}{H_\mathrm{g}}\right|^{-1}\Omega_\mathrm{K}^{-1},
\end{equation}
where $H_\mathrm{g}$ is the disk's gas pressure scale height, $R$ is the distance from the central star, $\beta_T$ is the exponent of the radial temperature structure $T\propto R^{\beta_T}$, $\gamma$ is the heat capacity ratio, $z$ is the distance from the disk midplane, and $\OmK$ is the Keplerian angular frequency.  \cite{Urpin2003} already showed that the VSI's growth rate decreases at thermal relaxation times beyond this limit, which has been recently demonstrated in numerical simulations by \cite{Klahr2023}.
At the same time, VSI creates mostly meridional gas flows, and its ability to vertically corrugate the dust layer has been demonstrated in various simulations \citep{Stoll2016, Lin2019, Flock2020, Schafer2020, Schafer2022, Fukuhara2023, Pfeil2023}.

\revII{It is, however, unclear whether the VSI evolves fast enough to avoid the effects of dust settling and maintains a dust layer that is thick enough to provide the necessary fast thermal relaxation times.}
If the grains are large and the cooling times are long, it would be possible that the VSI is not able to develop in the first place or is not able to counteract the settling of the grains.
\reportI{Work by \cite{Fukuhara2023}  and \cite{Fukuhara2024} has demonstrated that an equilibrium state might be possible in which the turbulent mixing of the VSI exactly counteracts sedimentation. Their studies are based on a semi-analytic model of the VSI turbulence and assume constant settling-mixing equilibrium.}
\revII{As radiative cooling, dust dynamics, and dust coagulation are interdependent processes, they would have to be accounted for in a single simulation. Constructing such a fully self-consistent numerical model of the VSI in protoplanetary disks is beyond the scope of this work. If, however, reasonable approximations are made, aspects of this interplay can be studied with much simpler techniques. Here, we present hydrodynamic simulations of protoplanetary disks including four dust fluids of evolving grain sizes in order to investigate the impact of the VSI-induced turbulence on the cooling times. We omitted a full treatment of dust coagulation and only evolved the grain size based on an analytic dust growth model \citep[similar to][]{Birnstiel2012}. Instead of including radiative transfer calculations to model the impact of the evolving dust distribution on the radiative cooling, we relaxed the temperature perturbations on a timescale calculated from the local dust size distribution. Although these methods are only approximations, they enable the first simulations of protoplanetary disks in which the timescale for the thermal accommodation of gas and dust is directly linked to the dynamics of the simulated dust fluids.}

\section{Methods}
\label{sec:Methods}
\subsection{Thermal relaxation times}
{In protoplanetary disks, thermal relaxation of temperature perturbations in the gas is mostly achieved via thermal accommodation with the dust.} We thus approximated
\begin{equation}
\label{eq:Cooling}
    t_\mathrm{thin}^{\text{NLTE}}\approx t_\mathrm{coll}^\mathrm{gas}=\frac{\gamma}{\gamma-1}\frac{1}{n_\mathrm{s}\Bar{v}_\mathrm{gas}\sigma_\mathrm{s}}\, ,
\end{equation}
where $t_\mathrm{coll}^\mathrm{gas}$ is the thermal accommodation timescale of the gas with the dust due to collisions \citep{Probstein1969, Burke1983, Barranco2018}.
The thermal accommodation timescale therefore depends on the mean molecular gas velocity $\Bar{v}_\mathrm{gas}$ and the Sauter mean radius of the dust size distribution \citep{Sauter1926}:
\begin{equation}
\label{eq:Sauter}
    a_\mathrm{s}=\frac{\int n(a)a^3\, \mathrm{d}a}{\int n(a)a^2\, \mathrm{d}a}\, ,
\end{equation}
{where $a$ refers to the grain size and $n(a)$ is the number density distribution of the particles.}
\rev{From this, the number density $n_\mathrm{s}=\rho_\mathrm{d}/(\nicefrac{4}{3}\pi \rho_\mathrm{m} a_\mathrm{s}^3)$ and the collisional cross-section $\sigma_\mathrm{s}=\pi a_\mathrm{s}^2$ follow. Here, $\rho_\mathrm{m}$ is the material density of the dust and $\rhod$ is the dust volume density.}
In most cases, the size distribution of dust in a protoplanetary disk can be \rev{approximated} as a truncated power law by following
\begin{equation}
    n(a) = \frac{n_\mathrm{tot}(p+1) }{a_\mathrm{max}^{p+1} - a_\mathrm{min}^{p+1}}a^{p}\, ,
\end{equation}
where $n_\mathrm{tot}$ is the total dust number density and $\amin$ and $\amax$ are the minimum and maximum grain sizes that truncate the distribution. 
\rev{Typical values of $p$ are in the range of $-4$ to $-2$, depending on the physical details of the grain collisions.} 

Considering a discretized version of the size distribution with bins of size $\Delta a=a_{i+\nicefrac{1}{2}} - a_{i-\nicefrac{1}{2}}$, we can write the same size distribution as
\begin{equation}
    n(a) = \sum_{i=1}^N   \frac{n_i\, (p+1) }{a_{i+\nicefrac{1}{2}}^{p+1} - a_{i-\nicefrac{1}{2}}^{p+1}}a_i^{p}\Theta(a_{i+\nicefrac{1}{2}}-a)\Theta(a - a_{i-\nicefrac{1}{2}}) \, ,
    \label{eq:DiscreteDistr}
\end{equation}
where each bin $i$ contains a total number density of $n_i$ and the size-grid cell interfaces are $a_{i-\nicefrac{1}{2}}>\amin$ and $a_{i+\nicefrac{1}{2}}<\amax$. The term $\Theta$ denotes the Heaviside step function and truncates every bin at $a_{i+\nicefrac{1}{2}}$ and $a_{i-\nicefrac{1}{2}}$. This assumes that the size distribution is a continuous power law with exponent $p$ in each bin.
With these definitions, we can rewrite the numerator of \cref{eq:Sauter} as
\begin{align}
    \int n(a)a^3\,  \mathrm{d}a 
    = \frac{p+1}{p+4} \sum_{i=1}^N n_i \frac{a_{i+\nicefrac{1}{2}}^{p+4} - a_{i-\nicefrac{1}{2}}^{p+4}}{a_{i+\nicefrac{1}{2}}^{p+1} - a_{i-\nicefrac{1}{2}}^{p+1}} \nonumber
    \coloneqq \sum_{i=1}^N n_i \chi_i \,.
\end{align}
Likewise, we can write the denominator of \cref{eq:Sauter} as
\begin{align}
    \int n(a)a^2\, \mathrm{d}a =  \frac{p+1}{p+3} \sum_{i=1}^N n_i \frac{a_{i+\nicefrac{1}{2}}^{p+3} - a_{i-\nicefrac{1}{2}}^{p+3}}{a_{i+\nicefrac{1}{2}}^{p+1} - a_{i-\nicefrac{1}{2}}^{p+1}}\nonumber 
    \coloneqq \sum_{i=1}^N n_i \xi_i\, .
\end{align}
If the bin interfaces are fixed in time and if we assume constant $p$ in every bin at all times, $\xi_i$ and $\chi_i$ are constants.
With these definitions, the Sauter mean of the entire size distribution can be written as
\begin{equation}
\label{eq:SauterDiscrete}
    a_\mathrm{s}=\left.\sum\limits_{i=1}^N n_i \chi_i \right/ \sum\limits_{i=1}^N n_i \xi_i \, .
\end{equation}

The calculation of the Sauter mean radius in a hydrodynamic simulation of dust and gas is therefore reduced to the calculation of the dust number density $n_i$ for the given dust size bins. 
Knowledge of the Sauter mean allows us to calculate $t_\mathrm{thin}^{\text{NLTE}}$ from \cref{eq:Cooling}. If gas and dust are treated through radiative transfer, as in \cite{Muley2023}, $t_\mathrm{thin}^{\text{NLTE}}$ can be used to dynamically relax the gas and dust temperatures with respect to each other.
In a simpler use case, $t_\mathrm{thin}^{\text{NLTE}}$ can be used as a dynamically evolving cooling timescale of the gas via
\begin{equation}
    T^{(n+1)} = T_\mathrm{eq} + (T^{(n)}-T_\mathrm{eq})\exp\left(-\frac{\Delta t}{t_\mathrm{thin}^{\text{NLTE}}}\right)\, ,
\end{equation}
where $T_\mathrm{eq}$ is the equilibrium temperature (set by stellar irradiation) and $\Delta t$ is the current simulation time step.
We demonstrate the latter case in hydrodynamic simulations of protoplanetary disks with a focus on the VSI activity in the next section.

\subsection{Hydrodynamic simulations}
We set up axisymmetric simulations in the $r-\vartheta$ plane of a protoplanetary disk in spherical coordinates. Our disk's initial hydrostatic structure follows the standard accretion disk with mass $M_\mathrm{disk}$ by \cite{LyndenBell1974}, with a truncated power law in column density with exponent $\beta_\Sigma$ and \rev{characteristic radius} $R_\mathrm{c}$:
\begin{equation}
    \Sigma(R)=(2+\beta_{\Sigma})\frac{M_{\text{disk}}}{2\pi R_\mathrm{c}^2} \left(\frac{R}{R_\mathrm{c}}\right)^{\beta_{\Sigma}}\exp{\left[-\left(\frac{R}{R_\mathrm{c}}\right)^{2+\beta_{\Sigma}}\right]} \, .
\end{equation}
From this, the vertical disk structure in $z$ follows via
\begin{equation}
   \rho = \rho_\mathrm{mid} \exp{\left[\left(\frac{H_\mathrm{g}}{R}\right)^{-2}\left(\frac{R}{\sqrt{R^2+z^2}} - 1\right)\right]} \, ,
\end{equation}
where we approximated the midplane dust density as $\rho_\mathrm{mid}\approx \nicefrac{\Sigma(r)}{\sqrt{2\pi} H_\mathrm{g}}$.
The angular frequency in hydrostatic equilibrium follows accordingly as 
\begin{align}
    \frac{\Omega^2(R,z)}{\Omega_\mathrm{K}^2} = &\left(\frac{H_\mathrm{g}}{R}\right)^2\left(\beta_T+\beta_\rho-(\beta_\Sigma + 2)\left(\frac{R}{R_\mathrm{c}}\right)^{\beta_\Sigma+2}\right) \nonumber \\ 
&- \frac{\beta_T R}{\sqrt{R^2+z^2}} + \beta_T + 1 \, ,
\end{align}
where $\beta_\rho$ is the power law exponent of the radial midplane density profile $\rho_\mathrm{mid}\propto R^{\beta_\rho}\exp\left[-\left(\nicefrac{R}{R_\mathrm{c}}\right)^{2+\beta_\Sigma}\right]$.

For the dust, we initialized the simulation with a \cite*{Mathis1977} (MRN) distribution with four dust fluids, according to \cref{eq:DiscreteDistr}. We conducted simulations with three different initial dust grain sizes $a_\mathrm{ini}=$\SIlist{1;10;100}{\micro \meter}. \revII{As the grains undergo coagulation, we let the maximum grain size evolve in time following} {the monodisperse growth rate $\dot{a}_\mathrm{max}=\amax \varepsilon_0\OmK (1-\nicefrac{\amax}{a_\mathrm{lim}})$ \citep{Kornet2001}, where $\varepsilon_0=\nicefrac{\Sigma_\mathrm{d,0}}{\Sigma_\mathrm{g,0}}$ is the initial dust-to-gas column density ratio. Here, we have multiplied the original growth rate by the factor $(1-\nicefrac{\amax}{a_\mathrm{lim}})$ to bound the growth to the fragmentation limit. The equation then has an analytic solution for a given initial grain size $a_\mathrm{ini}$}:
\begin{equation}
    \amax (t) = \frac{a_\mathrm{lim} a_\mathrm{ini}e^{t\, \varepsilon_0\, \OmK}}{a_\mathrm{lim}+a_\mathrm{ini}\left(e^{t\,\varepsilon_0\, \OmK} - 1\right)} \, ,
\end{equation}
where the limiting particle size $a_\mathrm{lim} = \min(a_\mathrm{frag}, a_\mathrm{frag\text{-}drift})$ is determined by the fragmentation velocity $v_\mathrm{frag}=\SI{500}{\centi\meter\per\second}$, the radial pressure gradient, and the level of turbulence, characterized by the $\alpha$ parameter \citep[see][for definitions]{Birnstiel2012}.
As our simulations were initialized without turbulence, we set this parameter to a very low value of $10^{-5}$.  The particle growth was thus limited by the drift-fragmentation limit \revII{($\sim \SI{1}{\centi \meter}$)} instead of the turbulent fragmentation limit.
We did not change the mass content of the four dust-size bins during the growth. Only the bin interfaces and the respective mass-averaged particle size of each bin changed due to growth. Growth therefore also influences the thermal accommodation times. As mass is shifted to larger sizes, fewer small grains are present, and thus \revII{dust-gas} collision rates decrease with time.

The four \rev{passive} dust fluids were advected using the same technique as in \cite{Pfeil2023} and thus evolved under the influence of settling and radial drift \rev{in the terminal velocity approximation}. The respective velocity components were calculated from the dust populations' mass-averaged particle size, where an MRN size distribution was assumed within each bin at all times. The bin boundaries were defined to be equidistant in log space.
 \rev{The size grid was then defined between $\amin$ as the lower boundary of the first cell and $\amax$ as the upper boundary of the last cell.}

\begin{figure}[ht]
    \centering
    \includegraphics[width=\columnwidth]{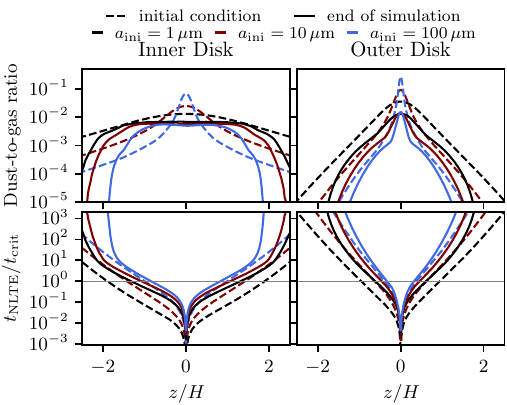}
    \caption[Averaged dust-to-gas ratios and thermal relaxation times in the inner and outer disk.]{Dust-to-gas ratios and thermal relaxation times radially averaged over the \SIrange{30}{55}{au} inner region and the \SIrange{115}{150}{au} outer region of our simulations. As the VSI develops, it keeps the dust particle at high altitudes and thus retains the necessary cooling times in the inner disk. The outer regions, however, are mostly sedimenting, as the VSI is not active there, except for the case with the smallest initial particle size.}
    \label{fig:InnerDisk}
\end{figure}

The initial dust distribution of each population followed the same vertical structure as the gas, but instead of the gas scale height with the dust scale height,
\begin{equation}
H_\mathrm{d} = H_\mathrm{g}\sqrt{\frac{\delta_\mathrm{ini}}{\delta_\mathrm{ini}+\St_i}}\, ,
\end{equation}
where $\St_i=\nicefrac{\pi a_i \rho_\mathrm{m}}{2 \Sigma_\mathrm{g}}$ is the Stokes number of the respective dust fluid and $\delta_\mathrm{ini}=\alpha$ is the initially assumed dust diffusivity.
Simulations initialized with different particle sizes also have dust layers of different initial heights and cooling times. This can be seen in \cref{fig:InnerDisk}, where the top panels show the initial distribution of dust in the inner and outer disk regions as dashed lines.

The cooling times increased with height since the dust density decreases with distance from the midplane. We show the cooling times in units of the local critical cooling time of the VSI. Values above one correspond to regions that are not susceptible to linear instability due to the vertical shear.

We ran the simulations for 1000 orbital timescales at \SI{50}{au}, that is, \SI{353553}{yr}. We present the simulation parameters in \cref{tab:params}.

\begin{table}
    \caption{Simulation parameters of the three presented runs. The only difference between the simulations is the initial particle size.}
    \centering
    \begin{tabular*}{\columnwidth}{@{\extracolsep{\fill}}lr}
        \toprule
        Simulation Parameter & Value \\
        \midrule\midrule
        Inner boundary $R_\mathrm{in}$ & \SI{25}{au} \\
        Outer boundary  $R_\mathrm{out}$ & \SI{150}{au} \\
        Vertical extend $\Delta \vartheta$ & $\pm 3 \nicefrac{H}{R}(\SI{50}{au})$ \\
        Resolution $N_r\times N_\vartheta$ & 2011$\times$512 \\
        \midrule
        Stellar mass $M_*$  & \SI{1}{M_{\odot}}  \\
        Disk gas mass $M_\mathrm{disk}$ & \SI{0.05}{M_{\odot}}  \\
        Temperature power law $\beta_T$ & -0.5  \\
        Column density power law $\beta_\Sigma$ & -0.85 \\
        Heat capacity ratio $\gamma$ & 1.4 \\
        Disk aspect ratio \nicefrac{H}{R}(\SI{50}{au}) & 0.07957  \\
        Disk \rev{characteristic} radius $R_\mathrm{c}$ & \SI{60}{au}  \\
        \midrule
        Minimum particle size $\amin$ & \SI{0.1}{\micro \meter}   \\
        Initial particle sizes $a_\mathrm{ini}$ & \SIlist{1;10;100}{\micro \meter}   \\
        Fragmentation velocity $v_\mathrm{frag}$ & \SI{500}{\centi \meter \per \second}   \\
        Initial dust-to-gas ratio $\varepsilon_0=\nicefrac{\Sigma_\mathrm{d,0}}{\Sigma_\mathrm{g,0}}$ & 0.01   \\
        Size distribution power law $p$ & -3.5   \\
        Initial dust diffusivity $\delta_\mathrm{ini}$ & $10^{-5}$   \\
        Dust material density $\rho_\mathrm{m}$ & \SI{1.67}{\gram \per \cubic \centi \meter}   \\
        \bottomrule
    \end{tabular*}
    \label{tab:params}
\end{table}

\section{Results}
\label{sec:Results}
\begin{figure*}[ht]
    \centering
    \includegraphics[width=\textwidth]{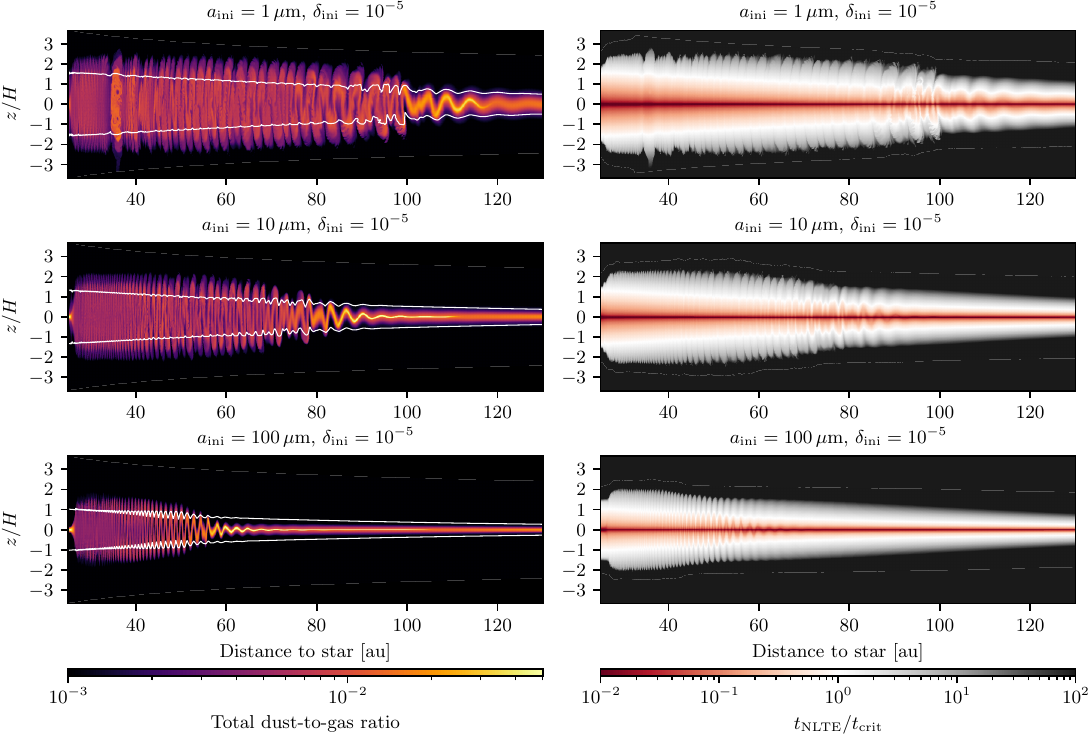}
    \caption[Dust-to-gas ratios and thermal relaxation times.]{Snapshots after $\sim$\SI{350000}{yr} of our protoplanetary disk simulations with thermal accommodation times \revII{calculated from the dust distribution for three different initial particle sizes}. Simulations initialized with larger particles and thus more settled dust layers are mostly not VSI active in the outer disk regions. If the VSI, however, can develop in the first place, it can stabilize the dust layer and sustain itself, even if grain growth is considered.}
    \label{fig:DtG_tNLTE}
\end{figure*}
We present the radially averaged vertical dust distributions and the respective cooling times at the beginning and at the end of the simulation in \cref{fig:InnerDisk}, where we distinguish between an inner disk region (\SIrange{30}{55}{au}) and an outer disk region (\SIrange{115}{150}{au}). In \cite{Pfeil2023}, we conducted simulations with static cooling times and showed that the VSI preferentially develops in the inner disks if the dust has already grown. We observed a qualitatively similar result here. The ongoing dust growth in the new simulations, however, modifies the results and imposes stricter conditions on whether the VSI can emerge. 
The turbulence is in every case stronger in the inner disk regions, where it maintains a relatively thick dust layer with \rev{an almost vertically constant dust-to-gas ratio} $\varepsilon$ and a stable vertical distribution of cooling times.
The outer regions show more settling due to the longer growth timescale of the VSI. Continued settling during the simulation is, however, of minor importance for the cooling times compared to the coagulation process because 
thermal relaxation is mostly done by small particles that have very long settling times. 

\rev{Whether} the VSI develops at all strongly depends on the initial maximum particle size and thus the initial \rev{vertical distribution of small dust grains}. If the particles are small at the beginning of the simulation, as in the case with $a_\mathrm{ini}=\SI{1}{\micro \meter}$, thermal accommodation times are short almost everywhere in the disk, and the VSI develops quickly. The resulting turbulence keeps the dust layer vertically extended and thus maintains the necessary cooling times self-sufficiently. This can also be seen in the first row of \cref{fig:DtG_tNLTE}, which shows the dust distribution and the respective cooling times at the end of the simulation. Regions inside of $\sim\SI{100}{au}$ show the typical filamentary VSI pattern in the dust densities and thus also in the cooling times. The outer regions develop slower, and the VSI is not present after \SI{350000}{yr} of evolution.

The situation is different in the simulation with initially large particles with $a_\mathrm{ini}=\SI{100}{\micro \meter}$. The particles are strongly settled at the beginning of the simulation, meaning that the cooling times in the upper disk atmosphere are accordingly long. 
The VSI begins to develop in the inner disk at a lower intensity than in the small-particle case. Nonetheless, a sufficiently thick layer of small dust is maintained, thus keeping the cooling times low enough in the inner regions.

The inner disk regions become VSI turbulent in all three cases, albeit on different timescales and at different intensities. 
The outer areas evolve differently. For small initial particle sizes, we observed that the VSI begins to develop at the end of the simulation. For larger particles, however, \rev{t}he outer disk areas are completely VSI inactive due to the already long cooling times at the beginning of the simulation. Regions beyond \SI{70}{au} are not VSI active for the largest initial particle size of $a_\mathrm{ini}=\SI{100}{\micro \meter}$. For ten times smaller particles, VSI is active up to \SI{90}{au} at the end of the simulation.

\begin{figure}[!t]
    \centering
    \includegraphics[width=\columnwidth]{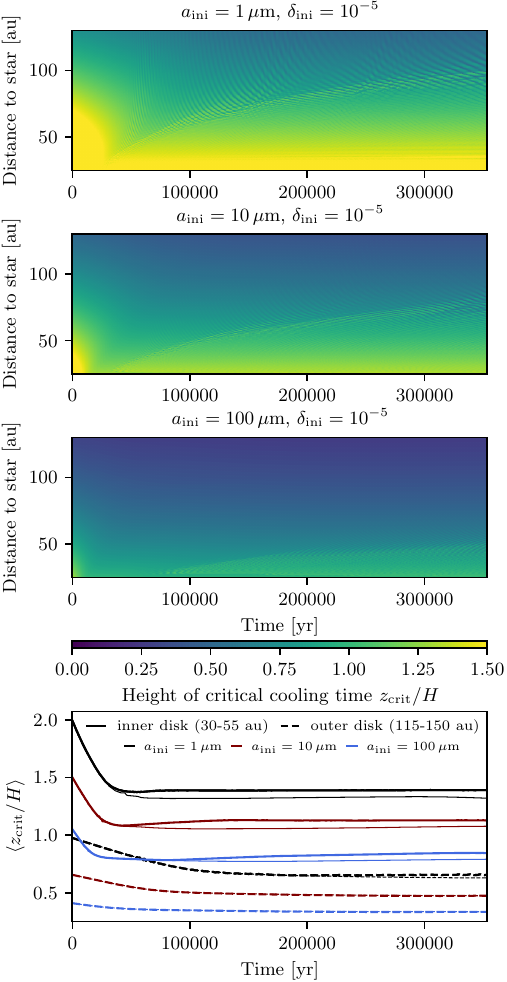}
    \caption[Evolution of the critical VSI height.]{Evolution of the vertical extend of the VSI susceptible region as a function of time and radius. In the inner disk, VSI quickly develops and stabilizes the settling dust layer in less than \SI{50000}{yr}. In the outer disk, VSI has not yet developed in our simulations. The thin lines show comparison simulations in which the VSI is artificially suppressed.}
    \label{fig:zCrit}
\end{figure}

To visualize the impact of the VSI on the cooling times, we have plotted the height at which the critical cooling time for the VSI is reached (\cref{eq:tcrit}) as a function of time and radius in \cref{fig:zCrit}. The beginning of all three simulations is characterized by the growth of the particles and the subsequent settling of the larger grains in the first $\sim \SI{40000}{yr}$. During this time, the surfaces of the VSI-susceptible zone move toward the midplane of the disk. This is mostly due to the coagulation process, \rev{which transforms small, not sedimented grains into larger and quickly sedimenting ones.} Cooling times are mostly determined by the small grains that remain in the atmosphere and only sediment slowly. The critical surface thus only moves toward the midplane because \revII{mass is transferred from small grains to big grains}. During this initial particle growth, the first VSI modes develop in the inner regions of the disk and soon begin to corrugate the dust layer. 
The emerging VSI turbulence stabilizes the extent of the VSI-susceptible zone and even moves small grains up into the disk atmosphere, thus extending the susceptible region with time in the inner disk.
In the simulation with the smallest initial particle size, the critical height reaches a stable value of $\sim 1.5\, H$ in the regions close to the star. If the simulations are initialized with larger, and thus strongly sedimented particles, we observed that the initial vertical extent of the susceptible zone is already small. For $a_\mathrm{ini}=\SI{10}{\micro \meter}$, the critical height is stabilized at $\sim 1.2\, H$; for $a_\mathrm{ini}=\SI{100}{\micro \meter}$, it is at $\sim 0.8\, H$.

In the outer regions, VSI either develops very slowly (in the case of the smallest initial particle size of \SI{1}{\micro \meter}) or not at all during the runtime of our simulations. 
This is a result of the limited simulation time and could also be a resolution issue, \rev{as the fastest growing modes become smaller with longer cooling times}. 
The trend is nonetheless clear: If a hydrodynamic simulation is initialized with large (\SI{100}{\micro \meter}), settled particles (here for $\delta_\mathrm{ini}=10^{-5}$), VSI only develops slowly, or possibly not at all, during the lifetime of a protoplanetary disk.
If the particles are initially small, or vertically dispersed, VSI can develop quickly and even extend the susceptible region by diffusing small particles to the upper disk layers.



\section{Discussion}
\label{sec:Discussion}
Our studies allow for a first insight into the interplay of VSI and its influence on the cooling times, and vice versa. There are nonetheless various limitations to the presented approach. 
We only modeled the size distribution with four dust fluids and assumed each size bin to have an MRN particle size distribution. At the beginning of the simulation, the size distribution is still continuous. When the dust fluids begin to drift and sediment, this changes, and the complete size distribution is no longer an MRN distribution. In the future, the coagulation and fragmentation process should be considered when the dust populations evolve. In that way, a meaningful size distribution could be maintained during the simulations.
\rev{We also set up the initial dust structure in settling-mixing equilibrium at all heights. This is strictly speaking only valid for Stokes numbers $\ll 1$ because the terminal velocity approximation is used in the derivations \citep{Dubrulle1995}. Thus, at large distances from the midplane, our initial conditions might not be realistic.
We modeled the dust as a passive fluid. Therefore, backreaction, which was shown to be able to quench the VSI activity in the disk midplane, especially if the dust settles, could not be accounted for \citep{Schafer2020}. The settling, however, occurs mostly once the dust has grown. We thus do not expect it to be able to hinder the VSI growth in the inner disk.

We have not tested other initial size distribution power laws than the MRN distribution. If initially more small dust were present, the conditions could be more favorable for the VSI. The same is true if the fragmentation velocity is smaller, which counteracts the production of large grains.

Furthermore, we have modeled the dust size as a vertically global quantity. Simulations of coagulation that consider a vertically varying dust size and coagulation at all heights have shown that dust growth can start in the upper layers, where the relative sedimentation velocities are high. This sedimentation-driven coagulation depletes the upper layers of grains, which coagulate, fragment, and sediment continuously \citep{Zsom2011}. Thus, a full radial-vertical treatment of coagulation must be applied in future studies.}

Other sources of cooling have also not been accounted for in our study. Especially in the upper disk atmosphere, cooling through molecular line emission could be the dominant channel of thermal relaxation \citep{Woitke2015}. These considerations, however, require thermo-chemical modeling of the different molecular species and are currently beyond the scope of the presented work.
A future self-consistent study of VSI with realistic thermal relaxation also has to involve the treatment of radiative transfer, ideally in a three-temperature framework, as recently presented by \cite{Muley2023}.
This is especially important when the effects of stellar irradiation and thermal relaxation in the optically thick regime should be included.

One of the caveats in the study of the VSI in protoplanetary disks remains the poorly constrained initial conditions of the simulations. Specifically, the state of the dust size distribution and the initial vertical extent of the dust layer are important for the onset of the VSI because they determine the cooling times.
If the dust at the beginning of the disk's lifetime is very small ($\sim$\SI{1}{\micro \meter}), we would expect the VSI to develop within the first \SI{350000}{yr} of disk evolution, even in the outer regions. Small dust would be expected if the grains resemble interstellar dust or if an intense source of turbulence during the disk formation process has caused strong fragmentation. The initial phases might be gravitoturbulent, which could cause such high initial levels of turbulence \citep{Gammie2001, Johnson2003, Hirose2017, Zier2023}.
Some observations and studies, however, hint toward early grain growth \citep{Galametz2019, Bate2022}. {Although these studies do not suggest the complete removal of sub-\si{\micro\meter}-sized grains, a reduction in the respective number density would already lead to less favorable conditions for the VSI, as shown by our simulations.}

\section{Conclusions}
\label{sec:Conclusions}
\revII{We have for the first time conducted hydrodynamic simulations of protoplanetary disks with thermal accommodation times that are consistent with the simulated dust densities and the present grain size.}
This made it possible to assess the influence of different initial conditions and dust grain growth on the developing VSI.
We find that the initial dust distribution has the biggest influence on the resulting spatial distribution of the turbulent gas flows. Coagulation also increases the cooling times and thus counteracts the VSI. This effect is, however, not a significant hurdle for the development of hydrodynamic turbulence in the inner regions of protoplanetary disks (inside of \SI{70}{au}), even if the dust is already large (\SI{100}{\micro \meter}) and vertically settled at the beginning of the simulation. 
The VSI can develop during the initial growth phase and even extend the susceptible region by diffusing small grains vertically. 
In the outer regions of protoplanetary disks (beyond \SI{70}{au}), the situation is more difficult for the VSI. If the dust is assumed to be large at the beginning of the simulation (\SI{100}{\micro \meter}) and is assumed to be in settling-mixing equilibrium (with a low initial diffusivity of $\delta=10^{-5}$), we find that the potentially VSI-susceptible region is \rev{constrained to a small area around the midplane} and will probably not develop \rev{VSI-induced} turbulence during the disk's lifetime.

\revII{We note that although our three simulations only differ in their initial dust grain size and reach identical maximum particle sizes, the resulting spatial distribution of dust and the level of VSI turbulence are vastly different between the three runs. Whether a specific disk stratification will develop VSI turbulence is thus highly dependent on the initial dust size distribution and the initial dust scale height}.

\begin{acknowledgements}
{The authors thank the anonymous referee for their helpful comments.}
T.P., H.K., and T.B. acknowledge the support of the German Science Foundation (DFG) priority program SPP 1992 “Exploring the Diversity of Extrasolar Planets” under grant Nos.\ BI 1816/7-2 and KL 1469/16-1/2. 
T.B. acknowledges funding from the European Research Council (ERC) under the European Union’s Horizon 2020 research and innovation programme under grant agreement No 714769 and funding by the Deutsche Forschungsgemeinschaft (DFG, German Research Foundation) under grants 361140270, 325594231, and Germany’s Excellence Strategy - EXC-2094 - 390783311.
Computations were conducted on the computational facilities of the University of Munich (LMU) and on the c2pap cluster at the Leibniz Rechenzentrum under project pn36ta.
\end{acknowledgements}

\bibliographystyle{aa} 
\bibliography{Literature}

\end{document}